\documentclass[journal, twocolumn]{IEEEtran}

\usepackage{graphicx}
\graphicspath{{./figures/}}
\DeclareGraphicsExtensions{.pdf}
\usepackage{epstopdf}
\usepackage{subcaption}
\usepackage{multirow}
\usepackage{amsmath}
\usepackage[numbers]{natbib}
\usepackage{url}

\begin{document}

\title{KINEMATIC ASPECTS OF RESIDUAL PRESTACK MIGRATION}

\author{Joerg F. Schneider \\
  Bureau of applied Geophysics, Nordstemmen, Germany}

\maketitle

\begin{abstract}

  In this contribution it is shown that various aspects of the concept of residual migration can be utilized for the case that a prestack time or depth migration has been performed for a seismic survey and a new depth is available for subsequent processing: The concept of residual migration is introduced by determining travel times of reflected events for individual traces from aplanatic curves. These events are migrated in suitable configurations for the new velocity model for which a raytracing set as required for a Kirchhoff prestack depth migration is available. The resulting transformations can be utilized in various ways to estimate important parameters such as the positional bias of  migrated events with respect to the corresponding zero offset position and the residual moveout for the new depth model; finally a new approach is suggested to estimate the aperture width  over the Fresnel zone for the first prestack migration. Various mapped prestack migrations can be estimated, including a prestack time migration with eliminated positional bias. The necessity of a new prestack depth migration can be assessed from the aperture width over the first order Fresnel zone and from the positional bias at a particular offset.  Additional kinematic parameters can be acquired for a residual prestack migration, in particular the curves over which the residual summation has to be performed. The approach is applicable to isotropic and anisotropic PSTM and PSDM. First applications are demonstrated with good results for a simple isotropic depth model with a laterally varying inhomogeneous velocity distribution with both isotropic and anisotropic PSTM and isotropic PSDM applications and for a vintage seismic survey. Computational costs of the implementation are low when compared to a full PSDM.\\
  In the appendix of the contribution a formula is derived for the migration velocities for a prestack time migration for small to intermediate source receiver offsets. The velocities can be determined either from an inhomogeneous anisotropic depth model or from residual moveout analyses of a previous migration; the application is demonstrated for an isotropic depth model.

\end{abstract}

\section{Introduction}

In the data processing of seismic surveys prestack migrations (PSM) have been performed for decades either as prestack time (PSTM) or prestack depth (PSDM) migrations (Yilmaz, 2001, Bednar, 2005, Robein 2010).\\
The task of migration and demigration and demigration of interpreted reflection responses has been considered for almost twenty years (Chauris et al., 2002, Lambare et al., 2007, Lambare et al., 2008,  Montel et al., 2009, Messud et al., 2017). The presented approach builds on this pioneering work; in the next section it will be attempted to indicate where differences exist to previous publications and where new aspects are considered. The transformation of first and second order reflection time parameters in PSTM and an extensive list of previous publications has been published by Iversen et al. (2012), see also Fomel (2014). Authors of a recent publication (Poliannikov and Malcolm, 2016) considered the case of the migration and demigration of interpreted reflection responses for the case of a PSDM for a variety of depth models. There are comparatively few publications which deal explicitly with solutions for residual prestack migrations (RPSM) (Rothman et al., 1985, Stolt, 1996, Etgen, 1998): in these publications various approximations to the wave equation were used to improve the quality of the seismic images partly with a transformation rather than a remigration of the original data; the computed results could be used in various ways, such as velocity analyses. In (Adler, 2002) a sequence of approximated residual migrations is estimated; the approach is applicable to inhomogeneous, anisotropic media for small velocity differences and is used for the optimization of migration velocities.\\
 Here a different strategy is pursued: individual horizons are considered for a migrated seismic survey, where a variety of parameters and mappings can be computed to assess the quality of a new depth model, the necessity of a full PSDM or to produce an estimate of the expected result: after PSM, migrated reflection responses are estimated from residual moveout (RMO) analyses. The migrated responses are demigrated and migrated into a new velocity model by applying aplanatic constructions: according to Hagedoorn (1954) the migrated response can be obtained as the envelope of all surfaces of equal reflection times of the source receiver pairs. For a new depth model a computed raytracing set is used to perform a raytracing migration, e.g. as a shot record migration (Reshef and Kosloff, 1986) or a modification for common offset sections. A novel extension is presented which opens the gate towards residual prestack migration:  from the response of a point diffractor in the mapped domain, the aplanatic curve is constructed over which the summation for RPSM has to be performed. The width of the aperture over the first Fresnel zone for the RPSM (Born, 1985,  Sheriff, 1980, Lindsey, 1989), is a further indicator for the necessity of a full PSDM. In the authors opinion the combination of the variety of the parameters and mappings offers new means for the evaluation of the migrated result. 

 \section{Method: Determination of aplanats and mapping}

 For the introduction of the concept of RPSM an outline of kinematic mapping will be presented: the upper figure in Fig. 1 shows an example of an interpreted result of a primary PSTM or Kirchhoff PSDM for a seismic survey with one migrated horizon $g_{0}(x,\tau/z)$ at zero offset, with lateral position x, migration time $\tau$ and depth z; S and G denote the position of a shot-receiver pair.  Spatially varying anisotropic parameters can be considered, for isotropic migrations according to the acoustic wave equation the migration velocities $v_{M} (x,\tau)$, viz. $v_{M} (x,z)$ are applied (Hubral and Krey, 1980). As an example a common offset migration for a particular offset h is considered: for an incorrect velocity model the migrated events do not align along $g_{0}$ but are observed at $g_{h}(x,\tau /z)$: in Fig. 1 $g_{h}$ is undermigrated  with RMO $P_{Th} P_{T0}$ at a particular location $x_{Th}$. In applications for seismic surveys the position of $g_{h}$ can be determined by RMO analyses (Schneider, 2011, Al-Chalabi, 2014), including higher order corrections where necessary.\\

 \begin{figure}%[hbt]
   \captionsetup{justification=justified, singlelinecheck=off}
\centering
\includegraphics[width=0.45\textwidth]{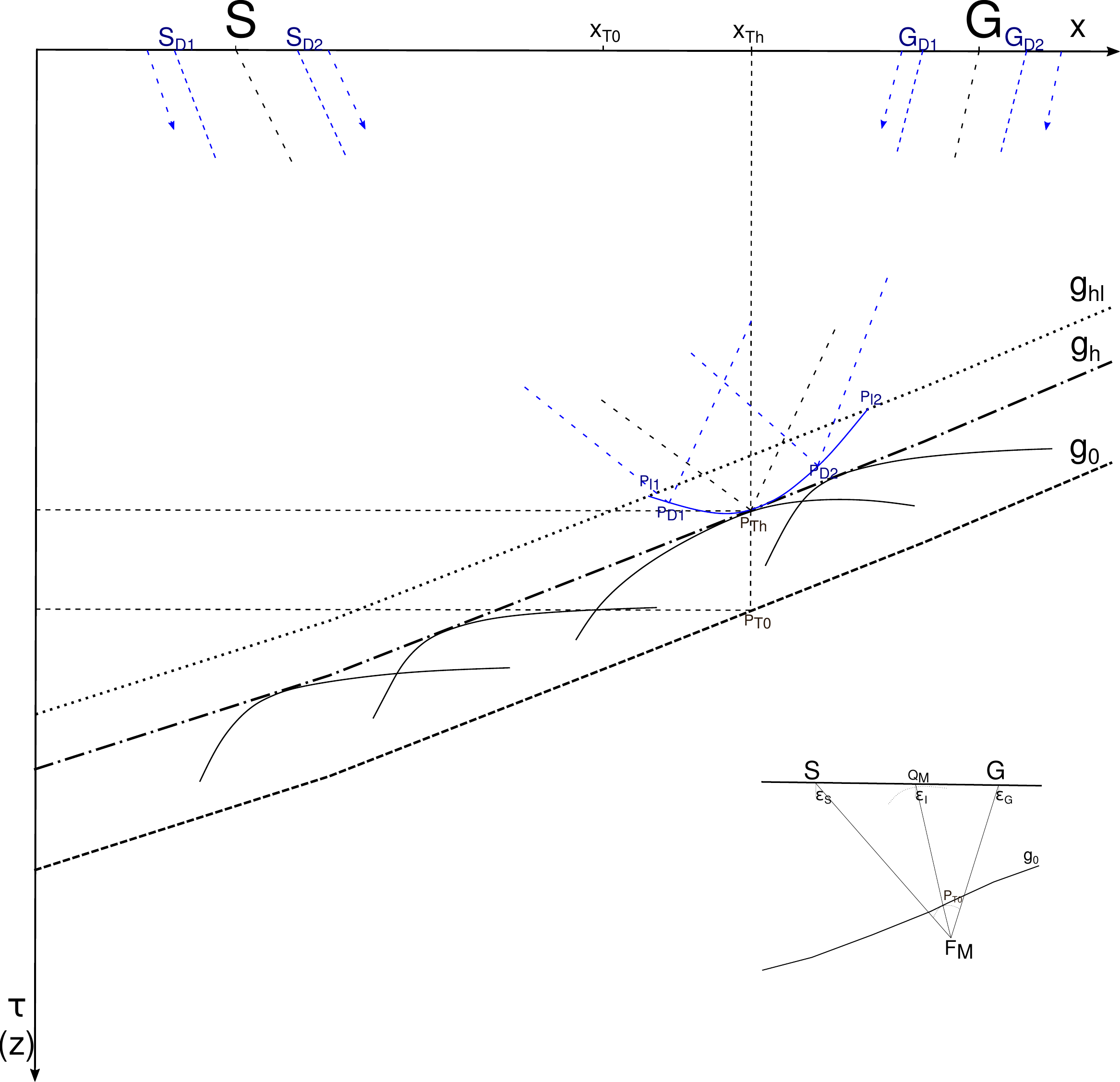}
\caption{PSTM/PSDM with aplanat construction for RPSM}
\label{fig:fig1}
\end{figure}

 \begin{figure}%[hbt]
   \captionsetup{justification=justified, singlelinecheck=off}
\centering
\includegraphics[width=0.45\textwidth]{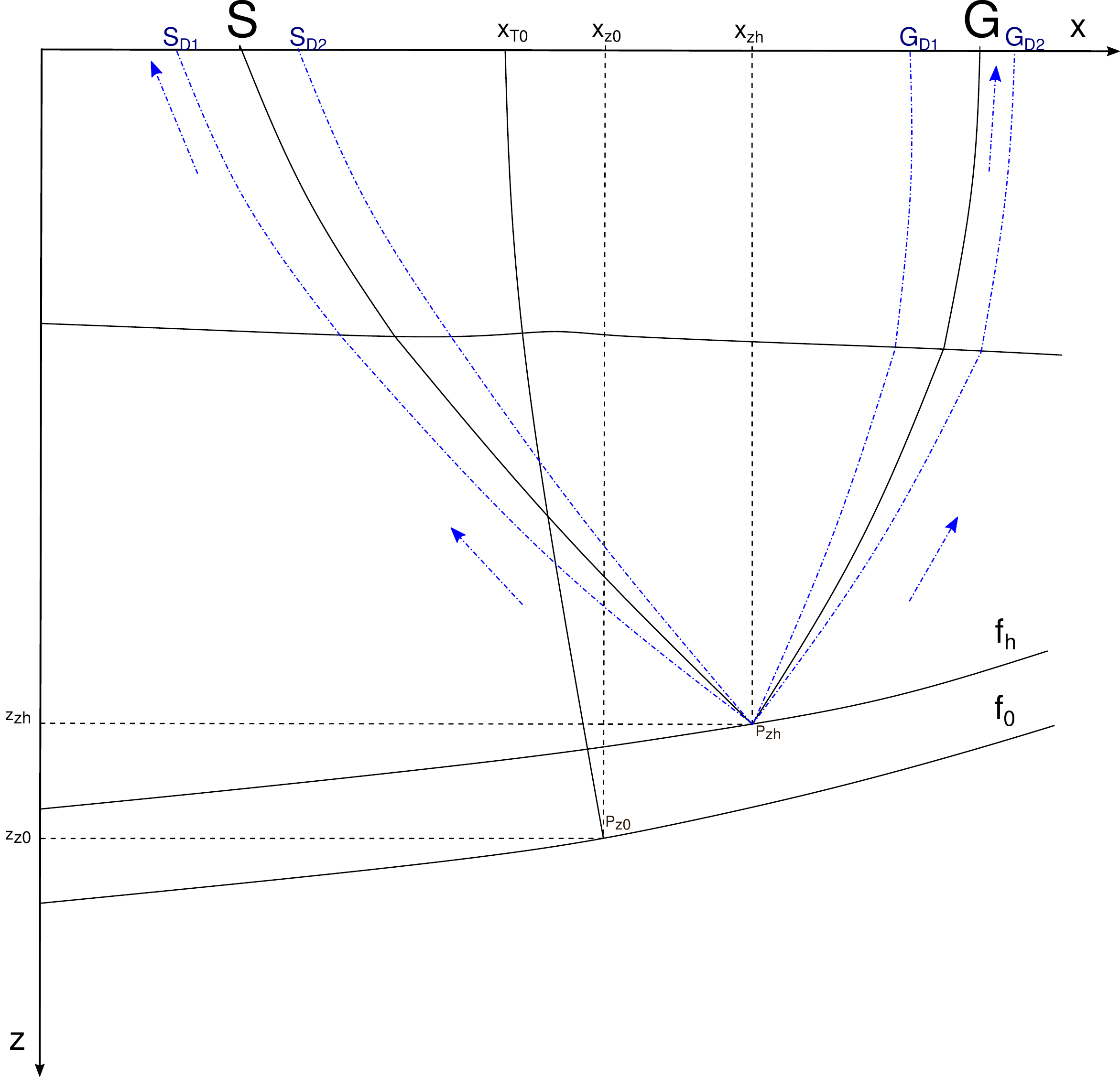}
\caption{Depth model for mapping and RPSM}
\label{fig:fig2}
\end{figure}

For simplicity it is assumed that at each source receiver pair only one arrival is observed from the considered reflector: $g_h$ is obtained by applying Hagedoorn’s principle with the aplanats shown in Fig. 1 (black) (in this context aplanats are considered as downward continued wavefronts at zero time). Aplanats have been considered by (Deregowski and Rocca, 1981, Hubral et al., 1996 and Bording and Lines, 1997). The proposed approach for RPSM is presented by considering aplanats from first principles; in the authors opinion it could have been formulated as an extension in terms of the terminology presented in (Hubral et al., 1995). It follows that the construction of the aplanat of constant traveltime with a common tangent at $g_h$ furnishes the traveltime $t_{SG}=t_S+t_G$ with the  traveltimes $t_S, t_G$ between source and receiver and $P_{Th}$, respectively. The construction of aplanats is achieved with the migration parameters (e.g. $v_M(x,\tau)$) for a PSTM or with the calculated raytracing set in the case of a Kirchhoff PSDM.  In Fig. 1 exactly one of these aplanatic curves has a point of tangency $P_{Th}$. $P_{T0}$ is the zero offset position on $g_0$ at $x_{Th}$. In this way reflection times for all shot receiver pairs are determined. It is possible to estimate spatial derivatives of reflection times in various configurations, e.g. for common shot/receiver gathers; in some applications care has to be taken that the derivatives are calculated with respect to the source/receiver coordinates at the surface (Iversen et al., 2012).\\
For the mapping it is now assumed that the new depth model has been rasterized in the form $v_F(x,z)$, e. g. by rasterizing an interface model as in Fig. 2.  For the individual traces of the record a series of migrated events are obtained; in Fig. 2 these events line up along lines $f_h$ (viz. $f_0$ at zero offset) where for each point on $f_h$ there corresponds a point in the original PSM (e.g. $P_{Th}$ to $P_{zh}$ ($x_{zh}$, $z_{zh}$)  in Fig. 1, 2); both $f_h$ and $f_0$  are supposed to be single valued with respect to their respective lateral coordinates. It should be noted that $f_h$ and $f_0$ do not coincide, i.e. the velocity model $v_F(x,z)$ is not necessarily true ($v_F$ in Fig. 2 will actually enhance the undermigration observed in Fig. 1). \\
For the RPSM aspect we consider $P_{zh}$ as a diffraction point: the constant offset response can be obtained in several ways in the vicinity of S and G, here for two diffracted rays in Fig. 1,2 in blue. It is then possible to invert the illustrated process of construction of aplanats and to obtain the aplanat from the diffracted rays at $P_{Th}$ in Fig. 1 (in blue), where two additional diffracted rays are shown. This aplanat defines the curve over which the RPSM summation has to be performed (Fig. 1, in blue, where $P_{D1}$, $P_{Th}$, $P_{D2}$ are points on the aplanat. The shape of the aplanat indicates that the original undermigration is enhanced). The described procedure is applicable to both anisotropic PSTM and PSDM; with the application of proper amplitude and phase corrections the summation will furnish the corresponding residual migration results (RPSTM and RPSDM, respectively) and should be performed over $x_{Fh}$, the aperture width over the order Fresnel zone (Born, M., 1985, Tabti et alii, 2004). For the amplitude and phase correction the heuristic approach suggested in (Schneider, 2019) can be employed. $x_{Fh}$ will be determined for the considered offset h: in Fig. 1 $x_{Fh}$ equals the distance $P_{l1}P_{l2}$ between the intersections of the aplanat with $g_{hl}$: in contrast to Tabti et alii (2004) where the time responses of $g_0$, $g_{h1}$ are separated by a distance of half a typical wavelength, here the aperture width of the post-migration Fresnel zone (Schneider, 1989, see also Tabti et al, 2004 for  similar applications) will be estimated; as shown by Sun (1996) both Fresnel zones are related . The most general method to determine the aperture width is the envelope construction described above: in Fig. 1 at $P_{Th}$ the position and slope are known; the radius of curvature and hence the shape of the aplanat for the RPSM can be determined by considering the common tangent of the diffraction aplanat (Fig. 1, in blue).  The computed mappings and values of $x_{Fh}$ are interpolated both laterally and vertically to provide a mapped PSDM or a RPSDM.\\
\\
Remarks:
\begin{itemize}
\item The reasoning in terms of Hagedoorn’s arguments is applicable to both PSTM and PSDM, for the isotropic and the anisotropic case. For the primary migration it can be applied to any PSTM/PSDM scheme for which aplanats can be constructed. It is the most general method to determine traveltimes and migrated events which can be applied with numerical efficiency to isotropic and anisotropic PSTM and PSDM. For the case that a mapping or RPSM is to be performed after an initial PSDM, different methods of migration and inverse migration can be used (Lambare et alii, 2008, Guillaume et alii, 2008). For the examples to be discussed below, several techniques have been employed.
\item In Fig. 2 at $P_{zh}(x_{zh}, z_{zh})$ the RMO $z_{zh}-z_{zh0}$ is observed, where $f_0$ is obtained from $g_0$ by zero offset migration with ordinate value $z_{zh0}$ at $x_{zh}$; it can be incorporated into the mapping and will also supply information about the validity of the model $v_F$ . Alternatively, if it is decided that a full PSDM is to be performed, the mapped parameters can be used either directly or as an excellent starting set for a subsequent RMO analysis. Similar computations are demonstrated in (Lambare et al., 2007), here the parameters are estimated and presented along individual horizons.
\item $P_{zh}$, $P_{z0}$ in Fig. 2 are the migrated offset and zero offset events for the gather at $x_{Th}$ in Fig. 1. The knowledge of the positional bias $x_{zh}-x_{z0}$ in Fig. 2 can be utilized for a transformation from $f_h$ to $g_0$ (Fig. 1,2) e.g. for a mapped PSTM with eliminated positional bias with respect to the depth model $v_F(x,z)$. This mapping can be performed as a one to one transformation; alternatively the application of a residual summation is possible, for which no examples are shown in this contribution.  After the mapping the false position of the reflector at zero offset can be corrected by well known applications, if necessary (Hubral, 1977, Whitcombe, 1994). The latter approach can be applied if the migrated survey is to be interpreted in time: if the velocity model $v_F$ exhibits lateral discontinuities, a vertical depth to time transformation of a PSDM is not possible (in such cases it might become a challenge to accomplish the mapping).
\item The positional bias $x_{zh}-x_{z0}$ as a function of offset is also a measure of the change in lateral resolution for the mapping and hence for the necessity of a further PSDM. To the authors knowledge this parameter has not been used before.
\item The width $x_{Fh}$ of the aperture over the first Fresnel zone is an important parameter for a RPSM: in most cases the summation for the RPSM can be confined to the first Fresnel zone; the size of $x_{Fh}$ measures the extent to which diffractions have been resolved by the primary PSM. So far it has not been used for interpretational purposes.
\item It is possible to produce supergathers by using the horizon information, with an aperture related to $x_{Fh}$ . An example will be presented in the next section.
\item The numerical implementation can be accomplished efficiently. The computational effort for the mapping is at least an order of magnitude lower when compared to a full PSDM; for most applications this will also be true for a full RPSM, mainly due to the smaller  migration aperture. 
\end{itemize}

\section{Applications}
Fig. 3 features the depth model of two horizons with a horizon of constant dip angle of $9.92^\circ$ which intersects the depth axis at a depth of 600m where the x-axis in Fig. 3 begins at x=400m in the depth model; the velocity distribution above this reflector is shown in Fig. 5 in red; this type of variation is not uncommon, e.g. over salt structures. A flat basement is situated at a depth of 3700m with a constant velocity of 3000m/s below the dipping horizon. A seismic survey was simulated by using ray tracing techniques (84 receiver/shot, max. offset: 4150m, 42-fold).\\
The first application features a PSTM for the dipping horizon: Fig. 4 shows the migrated stack after the application of RMO (the estimated zero offset position is indicated in blue). This application is supposed to illustrate the significance of the positional bias: the PSTM of the response of the dipping horizon in Fig. 3 was computed for two different velocity configurations; an initial PSTM was performed with migration velocities which were derived from zero offset considerations. These parameters are shown in magenta in Fig. 5 and furnish good results for intermediate offsets up to ~2400m (further details are discussed in the appendix). However, there is significant RMO (Fig. 10, magenta) if the full range of offsets is considered: the values in Fig. 7 were obtained by computing semblance coefficients along the migrated zero offset positions, as shown in Fig. 4 (blue). The minimum of the magenta values in Fig. 7 at $x=4500m$ correspond to 40ms RMO at offset h=4000m (Fig. 10). The RMO can be improved by applying an anisotropic PSTM: coefficients $v_{P0}$ for a PSTM with vertical transverse isotropy (VTI) (Douma and de Hoop, 2006) are shown in Fig. 5  (blue), the corresponding anellipticity coefficients in Fig. 6; these coefficients were determined by minimizing RMO in individual gathers with suitable optimization schemes. The semblance coefficients are almost optimal (Fig. 7, blue; for the entire survey the RMO at maximum offset is less than two samples). The positional bias, i.e. the lateral difference of migrated events in the true model with respect to the corresponding zero offset rays features different results: Fig. 8 shows this bias at h=4000m for the two migration results (magenta, blue). The bias is slightly smaller for the anisotropic PSTM but values of more than 200m are not acceptable. As an alternative a mapped PSDM with the true velocity model was computed from the isotropic PSTM after a higher order RMO analysis (Schneider, 2011) as described in the previous section. The corresponding stack response is observed at the true reflector position, the semblance coefficients (Fig. 7, grey) indicate that there is hardly any RMO in the gathers. In summary it can be stated that for this reflector the anisotropic and the moveout corrected isotropic PSTM will show equivalent results with poor lateral resolution. In contrast, the result obtained for the mapped PSDM is satisfactory.\\
The second application for this simple depth model features a different result: Fig. 9 shows a PSDM where the response of the base was migrated with an erroneous velocity of 2930m/s between the two reflectors  (with estimated zero offset position in red). Again, a higher order horizon RMO analyses was performed but here the subsequent mapping for the true velocity model was achieved with depth migrated gathers. The semblance coefficients of the initial PSDM (Fig. 7, green) indicate significant RMO at large offsets (Fig. 10, red).  There is no RMO in the mapped PSDM (Fig. 7, red) and the migrated stack focuses at the correct depth of 3700m. On the other hand, the positional bias (Fig. 8, red) does not exceed 50m at maximal offset. For this case it would not be necessary to perform a new PSDM: as far as lateral resolution is concerned the moveout corrected PSDM for the erroneous velocity model would give satisfactory results whereas the false position of the reflector in Fig. 9 could be corrected by a zero offset transformation of the migrated stack, if necessary (Hubral, 1977, Whitcombe, 1994).\\
\begin{figure}%[hbt]
  \captionsetup{justification=justified, singlelinecheck=off}
\centering
\includegraphics[width=0.45\textwidth]{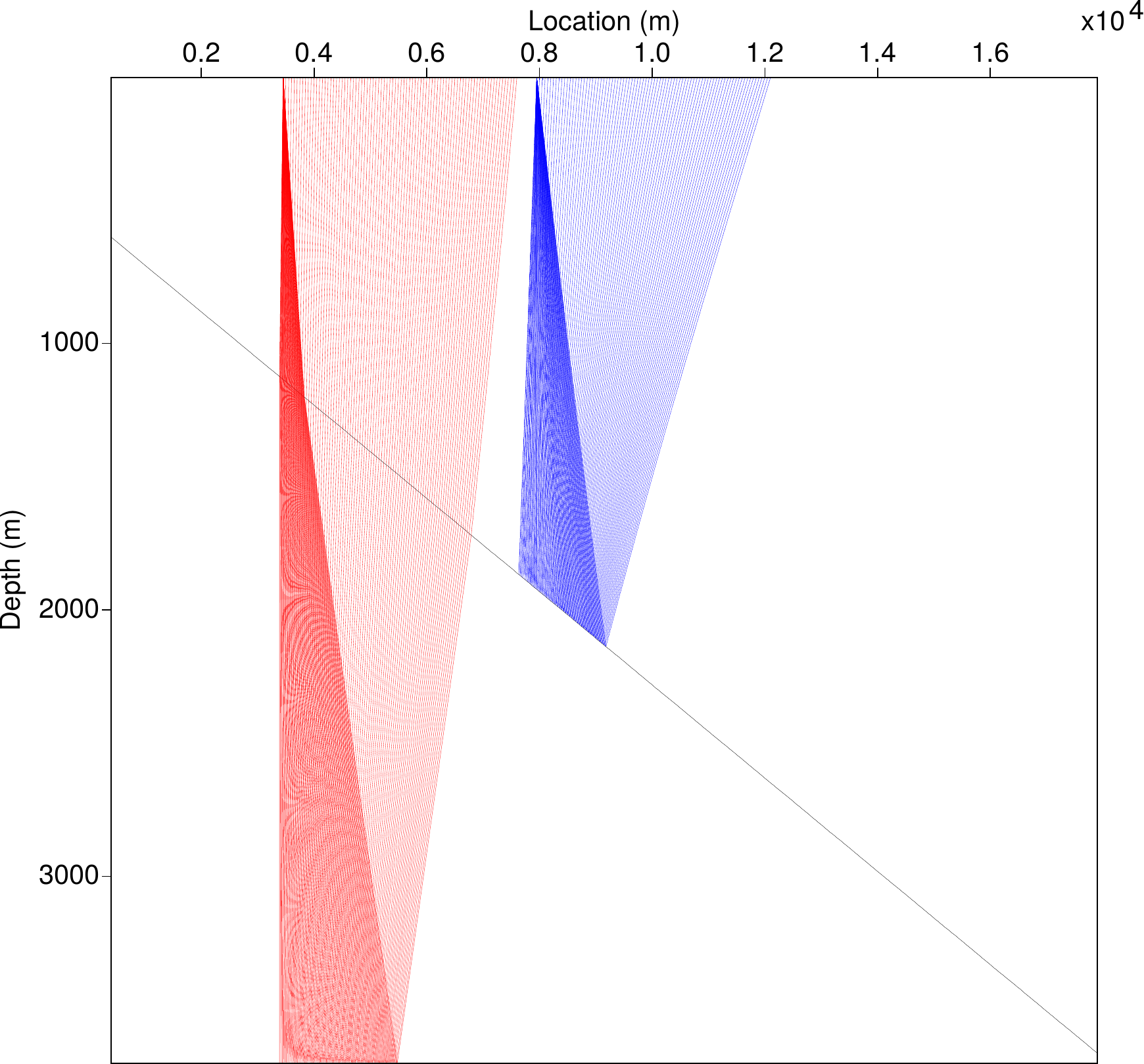}
\caption{Depth model with reflected rays for two horizons}
\label{fig:fig3}
\end{figure}

\begin{figure}%[hbt]
  \captionsetup{justification=justified, singlelinecheck=off}
\centering
\includegraphics[width=0.45\textwidth]{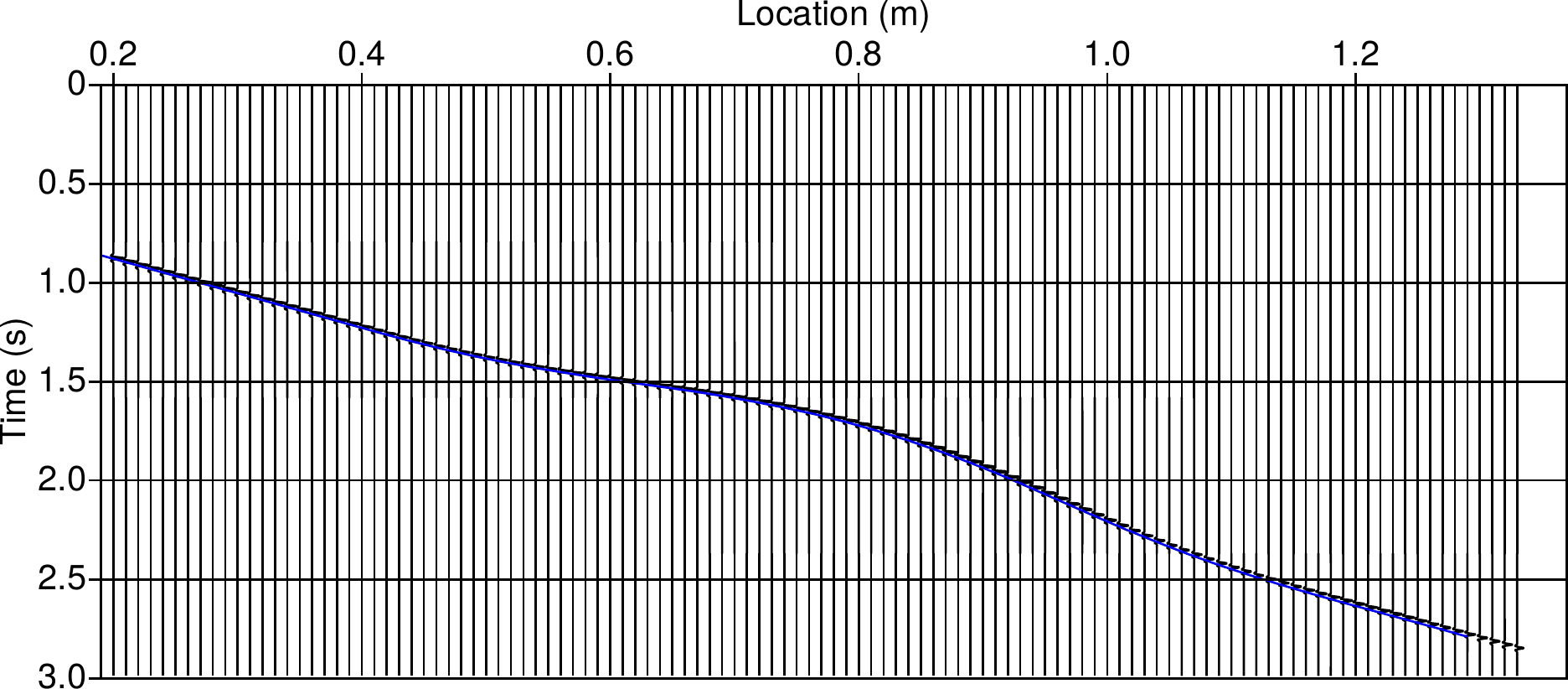}
\caption{Stacked PSTM with migrated reflector position(blue)}
\label{fig:fig4}
\end{figure}

\begin{figure}%[hbt]
\captionsetup{justification=justified, singlelinecheck=off}
\centering
\includegraphics[width=0.45\textwidth]{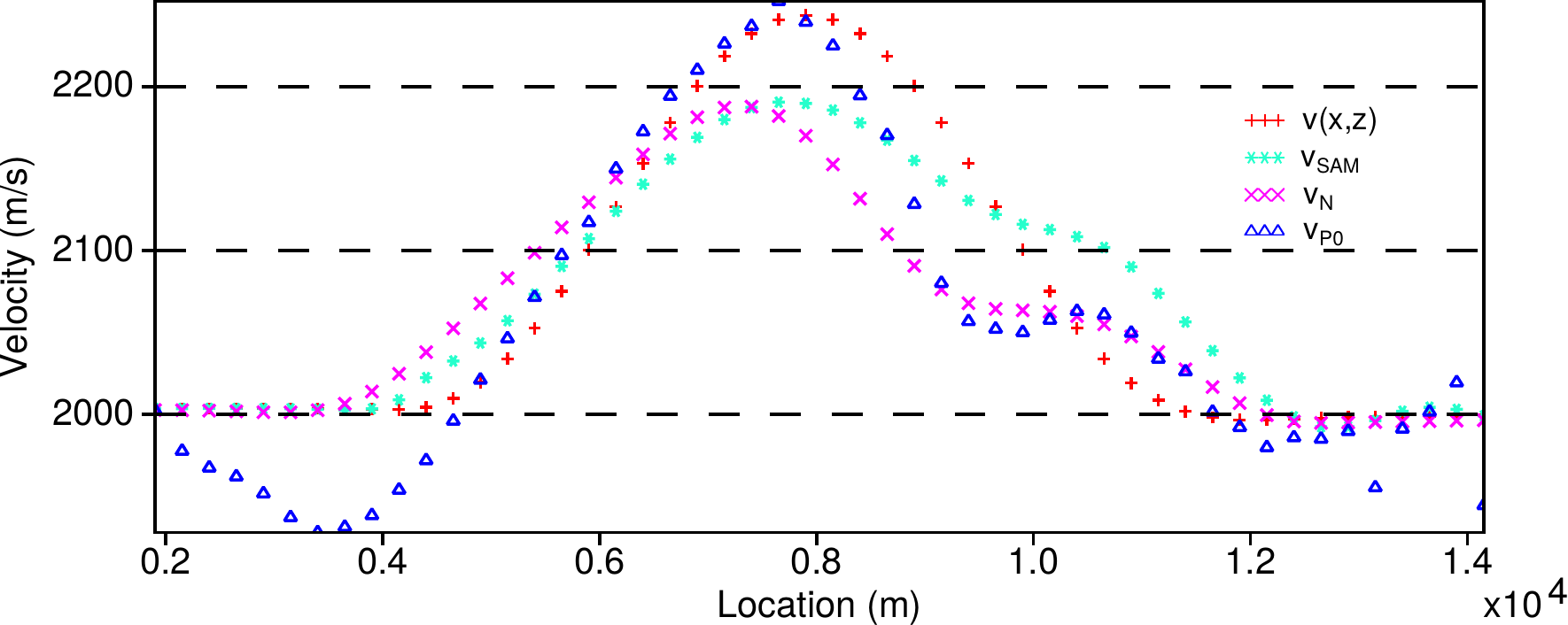}
\caption{Velocities for depth model and applications 1-3}
\label{fig:fig5}
\end{figure}

\begin{figure}%[hbt]
\captionsetup{justification=justified, singlelinecheck=off}
\centering
\includegraphics[width=0.45\textwidth]{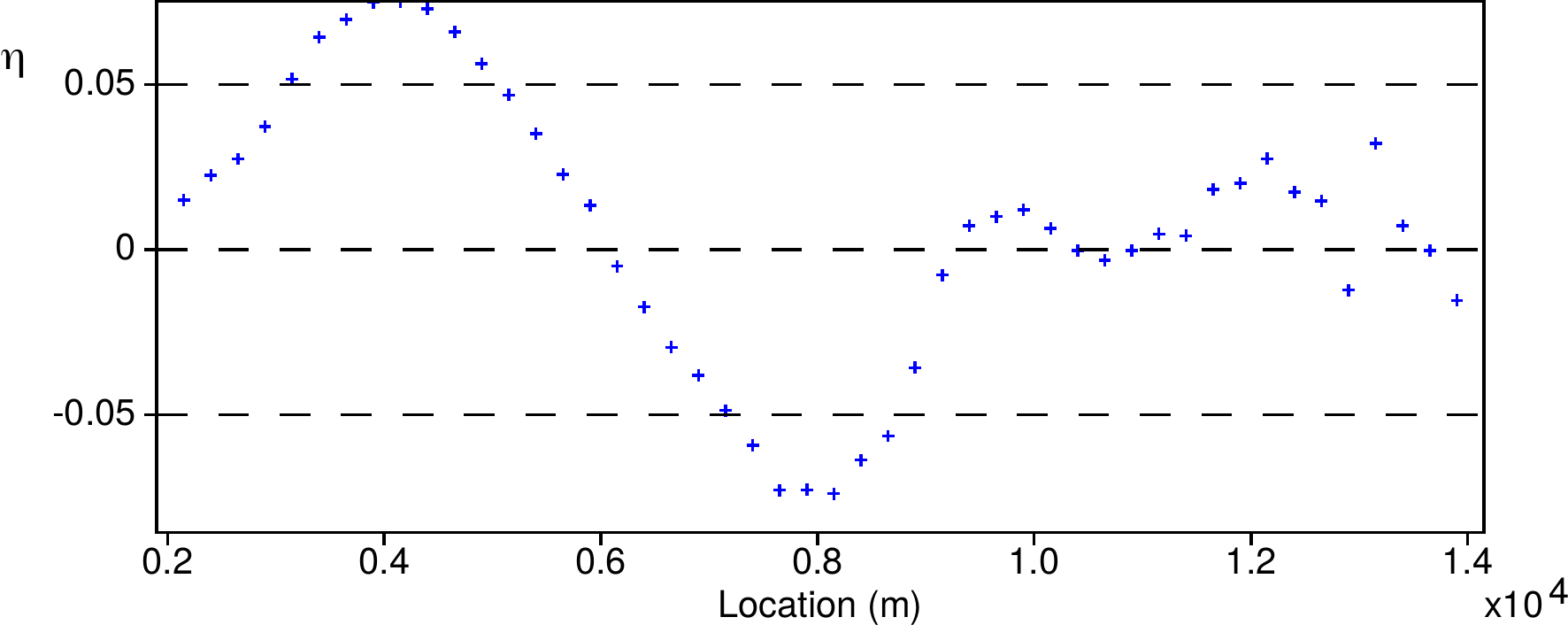}
\caption{Anellipticity coefficients for anisotropic PSTM}
\label{fig:fig6}
\end{figure}

\begin{figure}%[hbt]
  \captionsetup{justification=justified, singlelinecheck=off}
\centering
\includegraphics[width=0.45\textwidth]{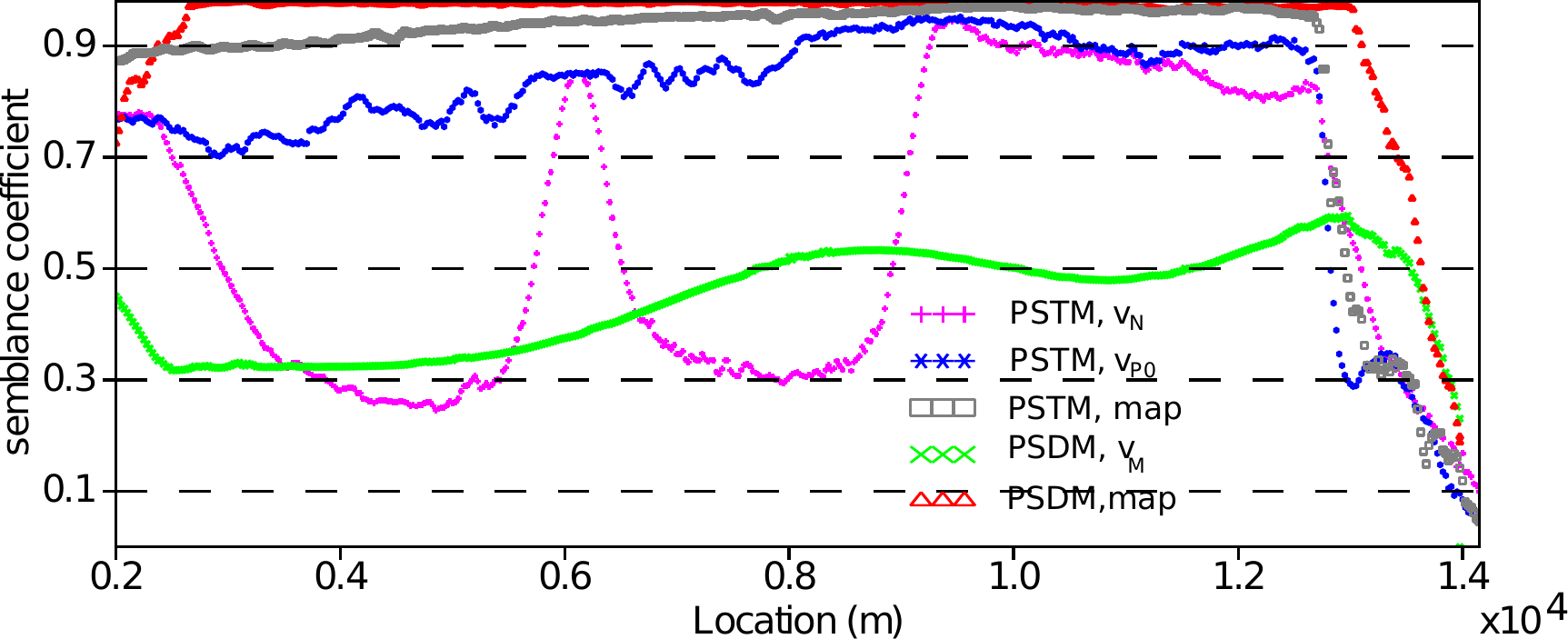}
\caption{Semblance coefficients for applications 1-3}
\label{fig:fig7}
\end{figure}

\begin{figure}%[hbt]
  \captionsetup{justification=justified, singlelinecheck=off}
\includegraphics[width=0.45\textwidth]{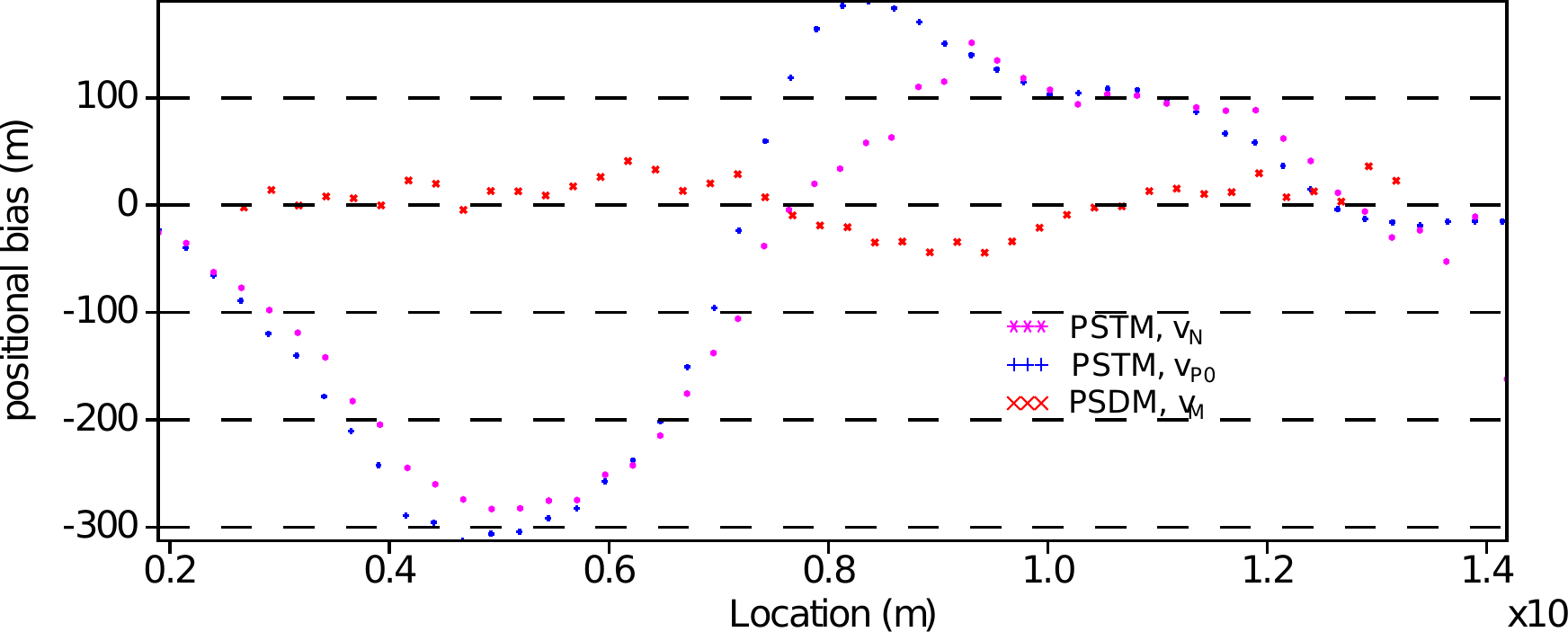}
\caption{Positional bias for applications 1,2}
\label{fig:fig8}
\end{figure}

\begin{figure}%[hbt]
\captionsetup{justification=justified, singlelinecheck=off}
\centering
\includegraphics[width=0.45\textwidth]{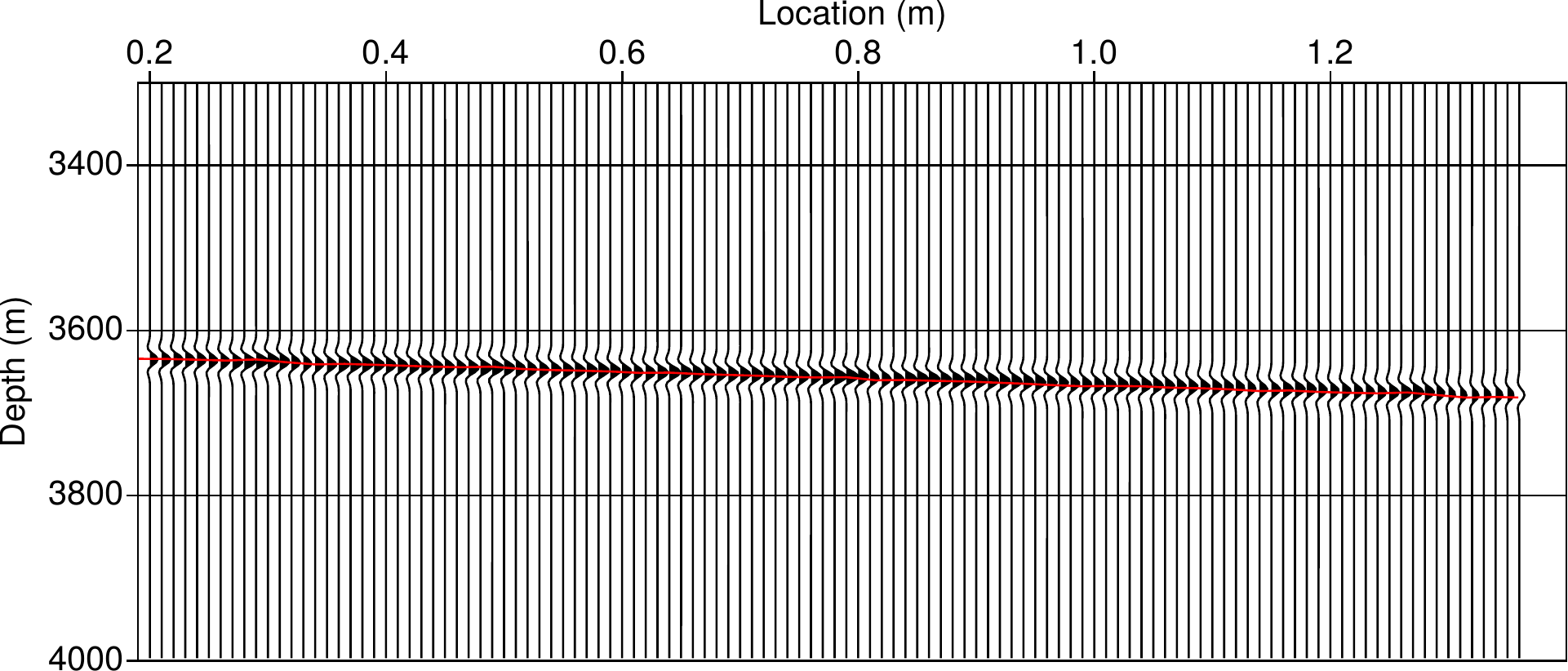}
\caption{Stacked PSDM with migrated reflector position (red)}
\label{fig:fig9}
\end{figure}

\begin{figure}%[hbt]
  \captionsetup{justification=justified, singlelinecheck=off}
\centering
\includegraphics[width=0.45\textwidth]{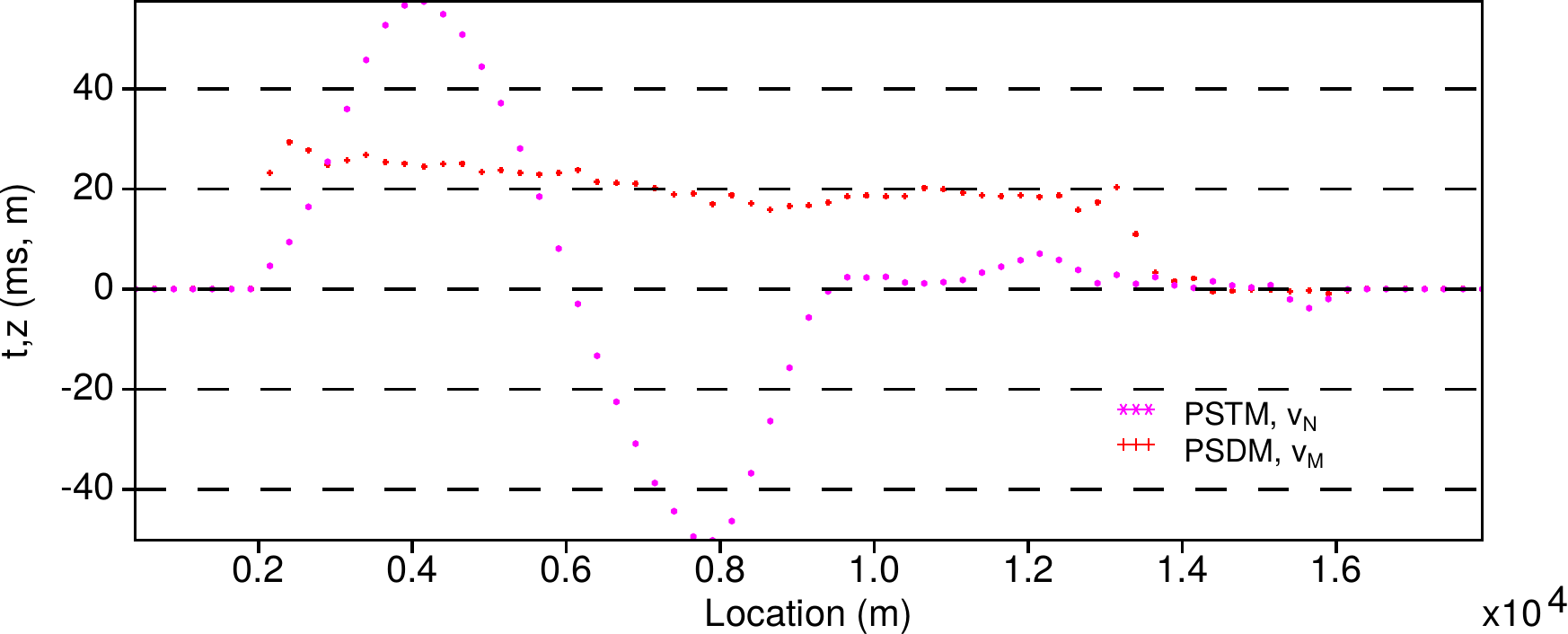}
\caption{RMO for applications 1,2 at maximum offset}
\label{fig:fig10}
\end{figure}

\begin{figure}%[hbt]
  \captionsetup{justification=justified, singlelinecheck=off}
\centering
\includegraphics[width=0.45\textwidth]{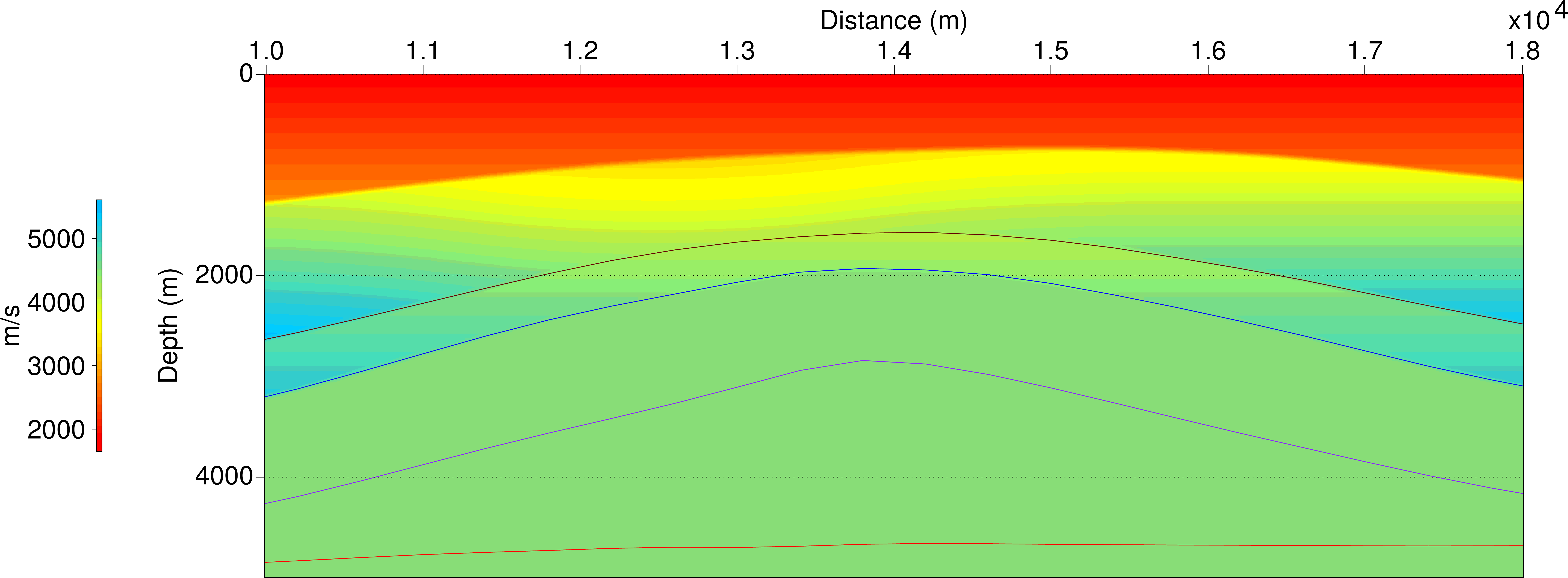}
\caption{Velocity model with four horizons}
\label{fig:fig11}
\end{figure}

\begin{figure}%[hbt]
  \captionsetup{justification=justified, singlelinecheck=off}
\centering
\includegraphics[width=0.45\textwidth]{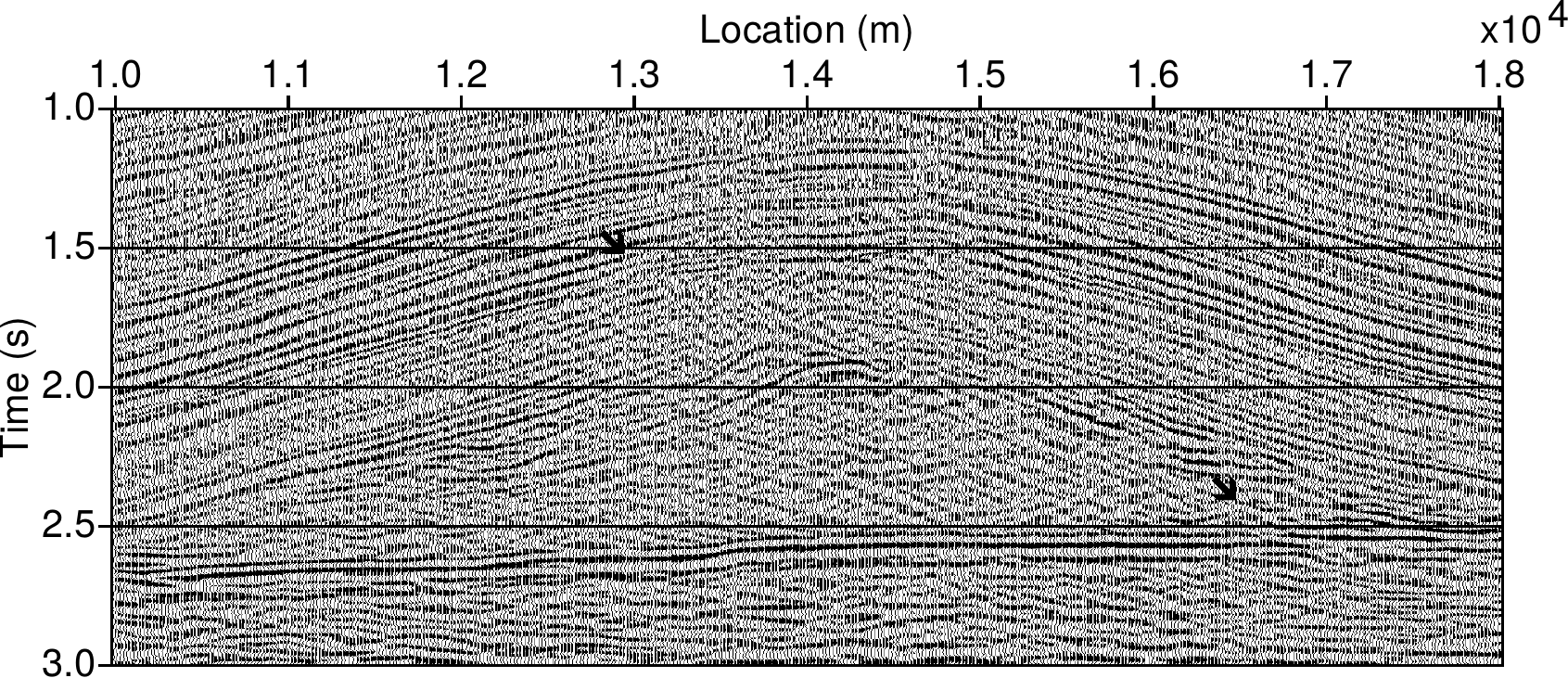}
\caption{PSTM after RMO}
\label{fig:fig12}
\end{figure}

\begin{figure}%[hbt]
  \captionsetup{justification=justified, singlelinecheck=off}
\centering
\includegraphics[width=0.45\textwidth]{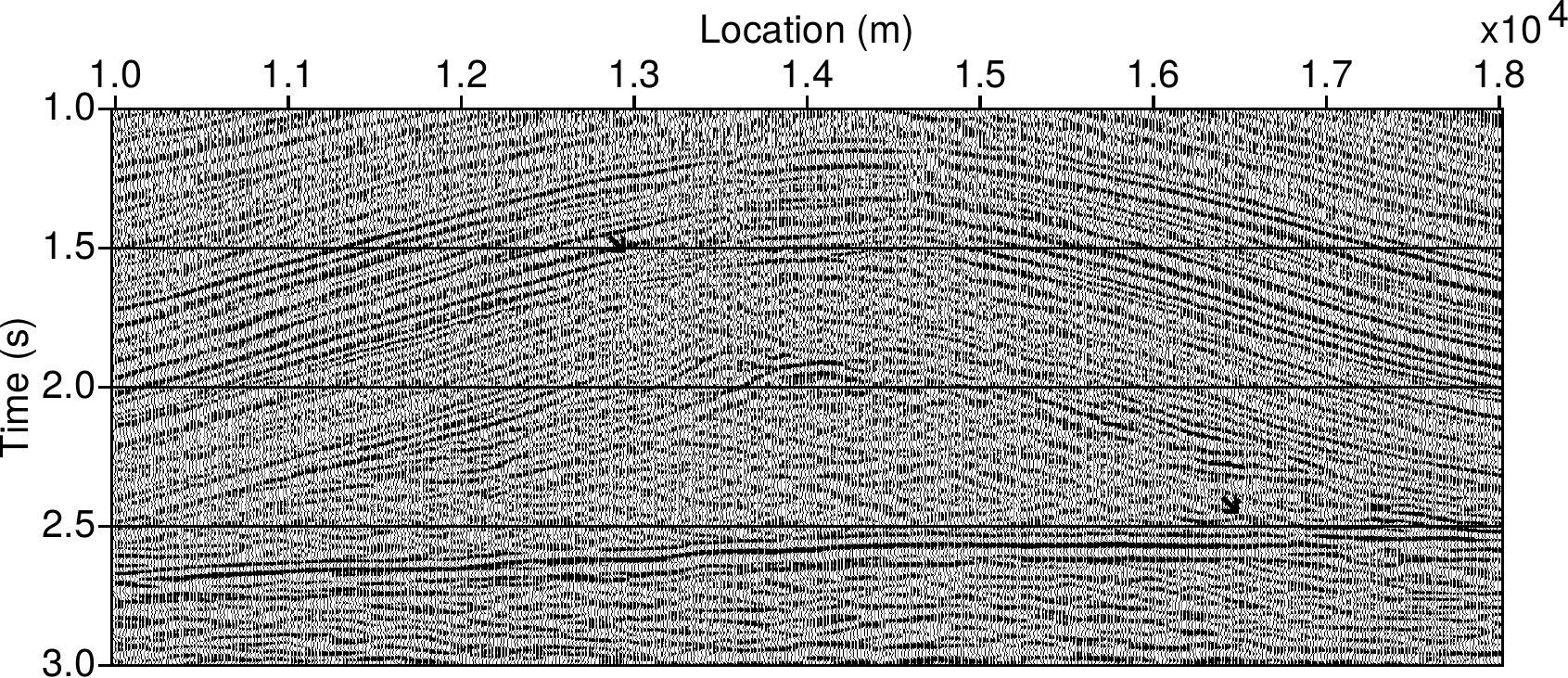}
\caption{Mapped PSTM}
\label{fig:fig13}
\end{figure}

\begin{figure}%[hbt]
  \captionsetup{justification=justified, singlelinecheck=off}
\centering
\includegraphics[width=0.45\textwidth]{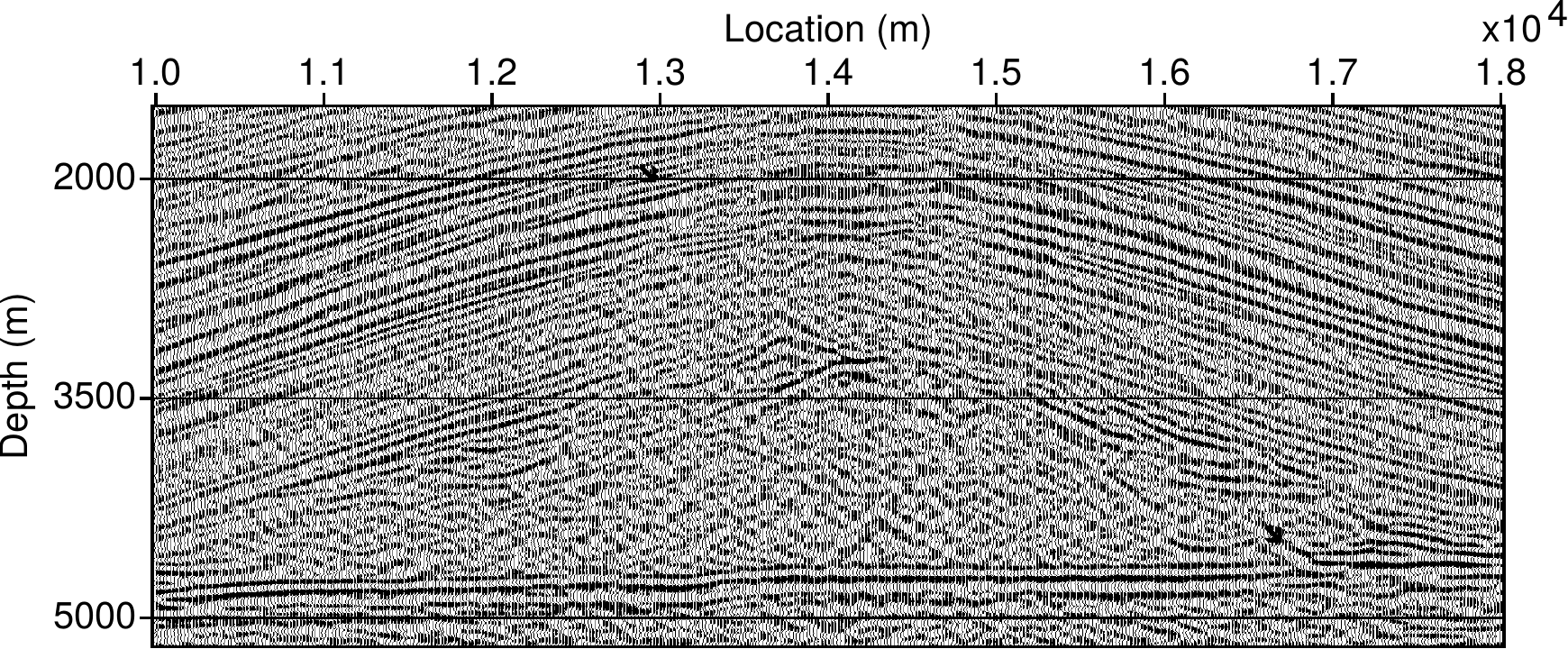}
\caption{PSDM after RMO}
\label{fig:fig14}
\end{figure}

\begin{figure}%[hbt]
  \captionsetup{justification=justified, singlelinecheck=off}
\centering
\includegraphics[width=0.45\textwidth]{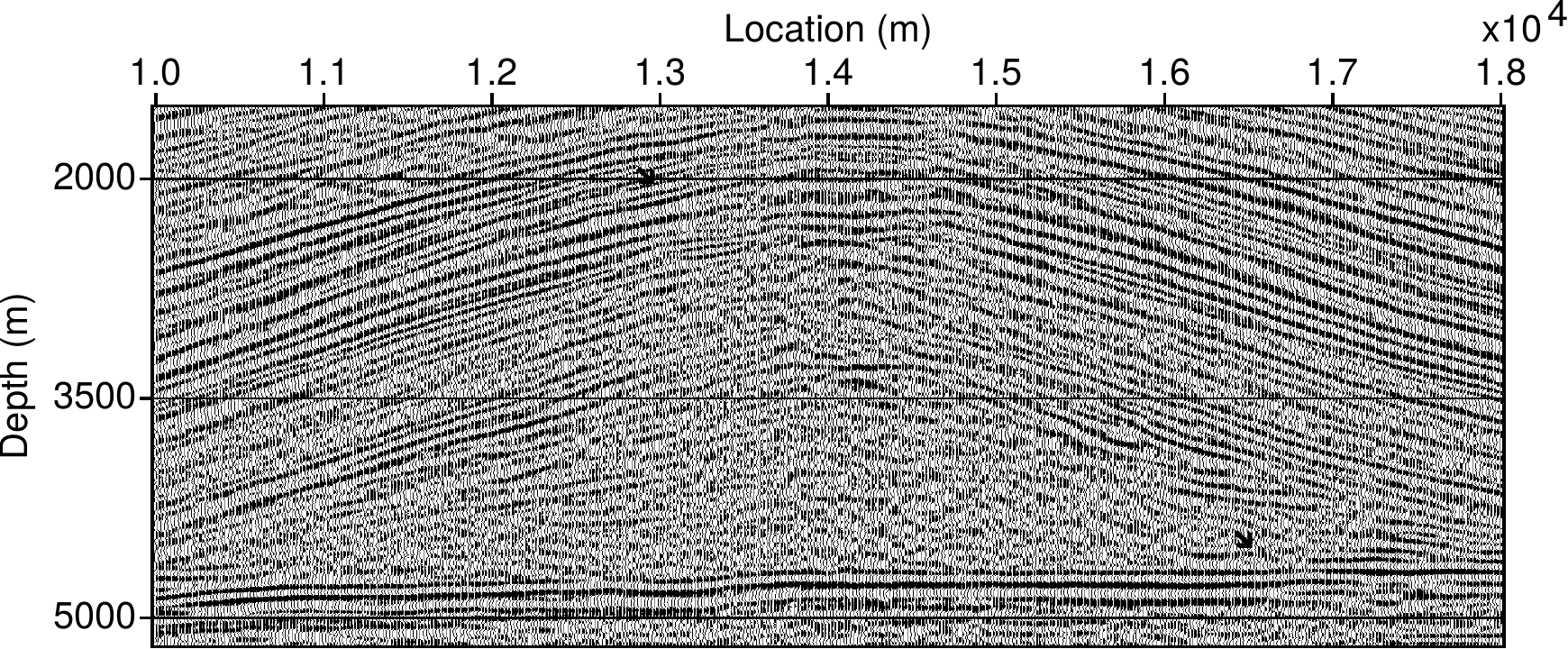}
\caption{Mapped PSDM}
\label{fig:fig15}
\end{figure}

\begin{figure}%[hbt]
  \captionsetup{justification=justified, singlelinecheck=off}
\centering
\includegraphics[width=0.45\textwidth]{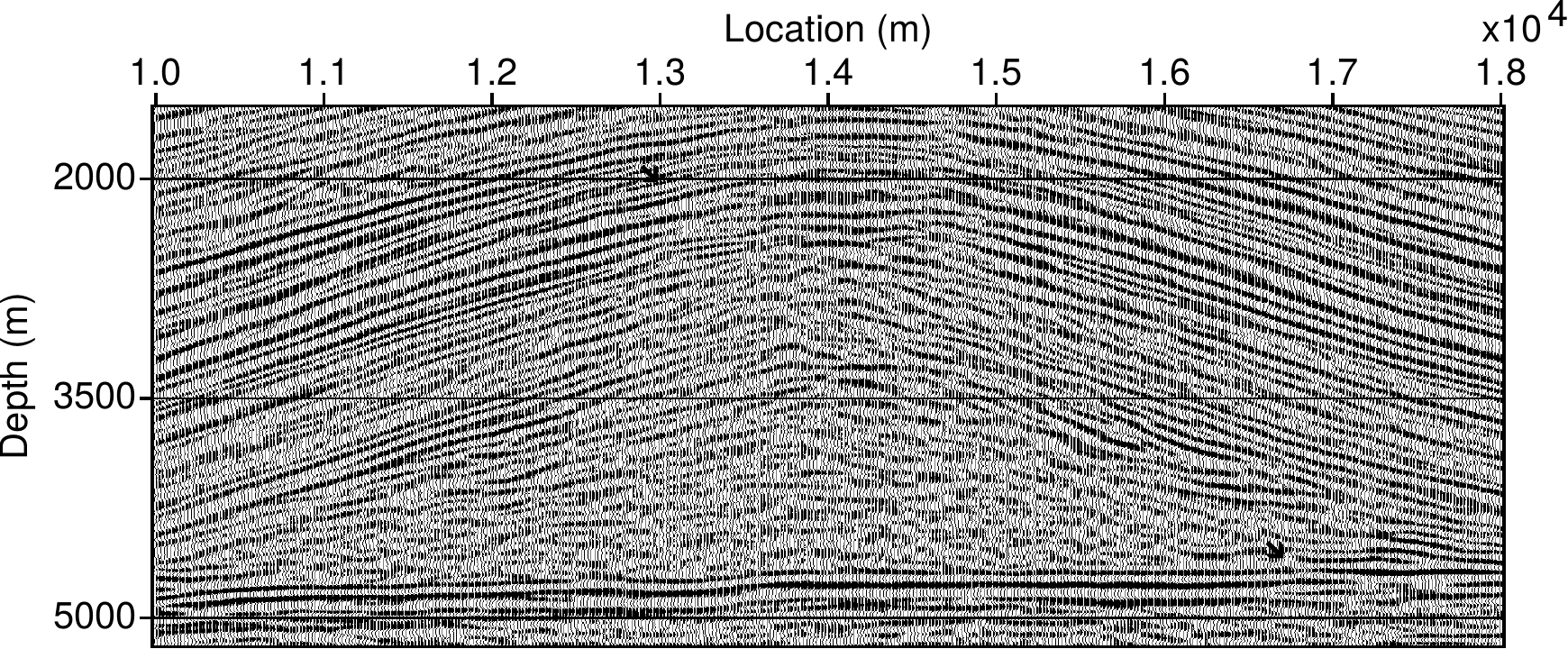}
\caption{Mapped PSDM (summation over 5 CIG)}
\label{fig:fig16}
\end{figure}

\begin{figure}%[hbt]
  \captionsetup{justification=justified, singlelinecheck=off}
\centering
\includegraphics[width=0.45\textwidth]{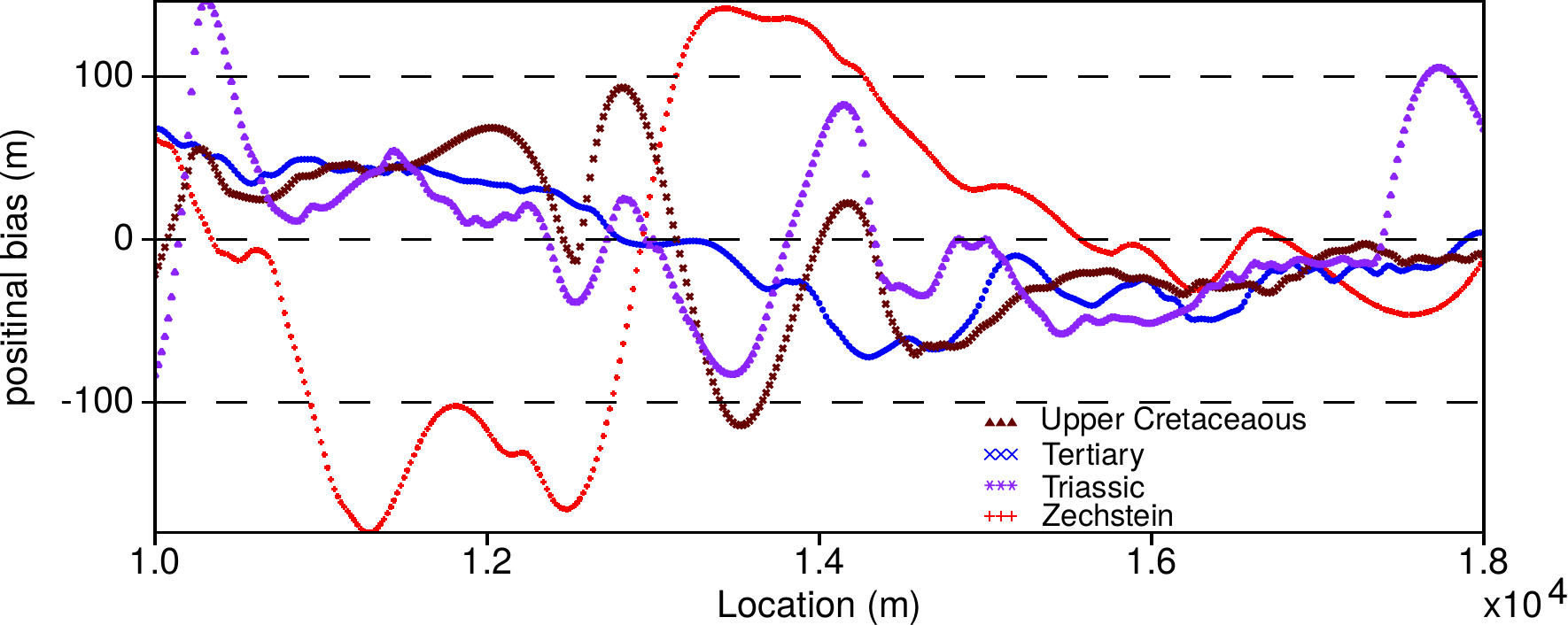}
\caption{Positional bias for PSTM versus mapped PSDM}
\label{fig:fig17}
\end{figure}

\begin{figure}%[hbt]
  \captionsetup{justification=justified, singlelinecheck=off}
\centering
\includegraphics[width=0.45\textwidth]{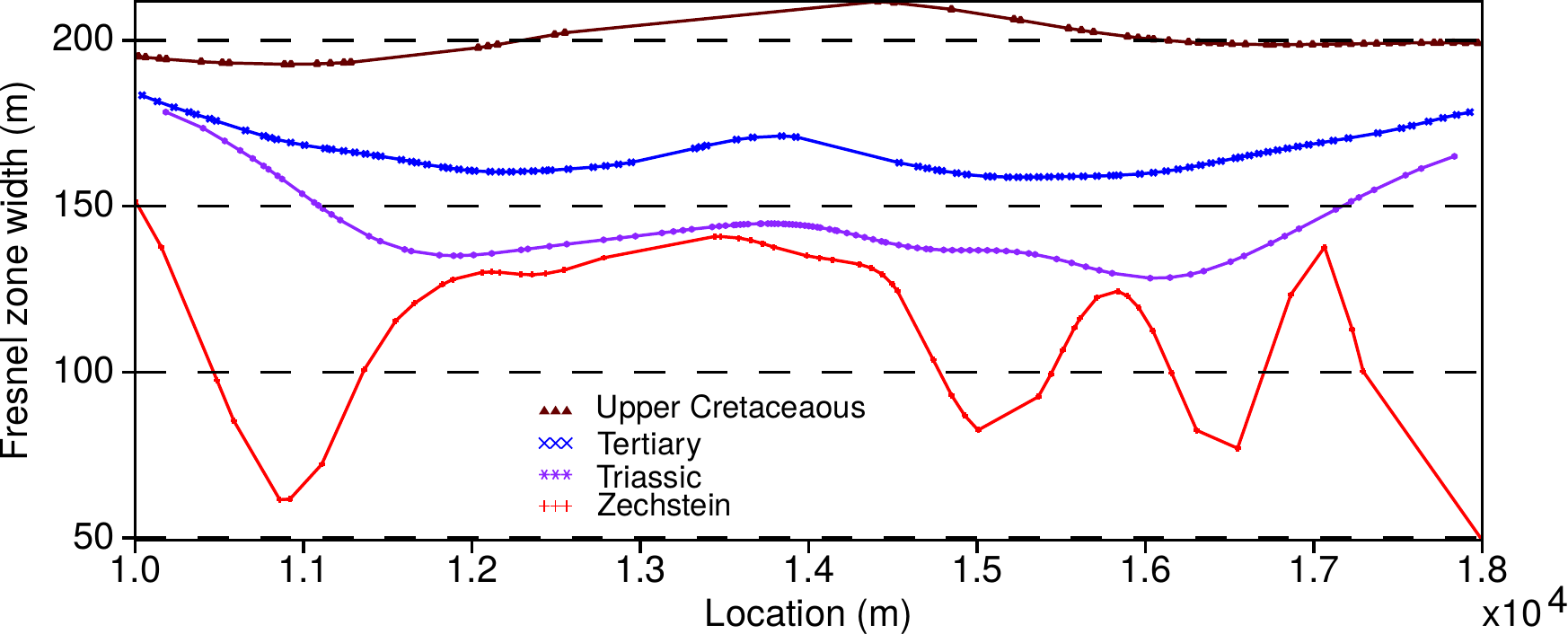}
\caption{Width of first Fresnel zones at maximum offset}
\label{fig:fig18}
\end{figure}

Fig. 11 shows a depth model with four main horizons over a moderate salt structure in Northern Germany. A vintage survey (maximum offset: 2400m, receiver spacing: 40m, 15-fold) was available for which both isotropic PSTM and PSDM were computed as the first migration. For the PSTM migration in Fig. 12 migration velocities were estimated along image rays in the depth velocity model in Fig. 11 (Hubral and Krey, 1980), the mapped section with eliminated positional bias is shown in Fig. 13. In the center of the section there are some migration artifacts above the base of Zechstein. The simple depth model was chosen to illustrate the applicability of the approach: differences are not too significant; some differences are marked in Fig. 12-15. The positional bias at maximum offset (Fig. 17) increases with depth and exceeds the separation of five trace spacing for the base horizon. Finally a PSDM application was considered: Fig. 14 shows the stack of the first PSDM, where the RMO analyses were again performed for the four horizons in Fig. 11. The velocity model for this migration features a constant negative bias of 80m/s for the upper cretaceous formation above the brown reflector. RMO corrections were included for the mapped PSDM in Fig. 15 which was computed with the original upper cretaceous velocity. Several differences exist, including a vertical shift with respect to Fig. 14 due the higher migration velocity. Fig. 16 shows an example of the formation of supergathers; in some cases the apparent increase in signal to noise ratio can help to identify structural details. The width of the first Fresnel zones in Fig. 18 decreases with the depth of the reflecting horizons where there are signs of instabilities for the base horizon. The values of the width of the first Fresnel zone are less than seven trace spacing for the basement.

\section{Conclusion}

The case is considered that RMO has been determined from horizon analyses after a prestack migration of a seismic survey. A new approach has been introduced, its application allows the application of important parameters: the positional bias of reflected events with respect to corresponding zero offset rays, the RMO to be expected for the new velocity model and the width of Fresnel zones for the original PSM. It is possible to estimate migrated sections for a new velocity model: from aplanatic curves the traveltimes of reflected events are determined for individual traces. These events are migrated in suitable configurations for a new velocity model for which a raytracing set as required for a Kirchhoff PSDM is available. It has been shown that the resulting transformations can be utilized in various ways: the validity of the new depth model can be assessed from the mapped residual moveout which can also be used for a subsequent migration. Various mapped prestack migrations can be estimated, including a prestack time migration with eliminated positional bias. It will be convenient to use the latter residual migration if an interpretation of the migrated section in time is desired and a vertical time to depth transformation of the depth migrated traces is not possible. Novel techniques have been introduced which can be used to assess the necessity of a new prestack migration from the aperture width over the first Fresnel zone and from the positional bias at a particular offset.  Kinematic parameters can be acquired for a residual prestack migration, including the curves over which the residual summation has to be performed. The approach is applicable to isotropic and anisotropic PSTM and PSDM. It has been demonstrated with good results for an isotropic depth model with a laterally varying inhomogeneous velocity distribution with both isotropic and anisotropic PSTM and isotropic PSDM applications and for a seismic survey. Computational costs of the implementation are low when compared to a full PSDM, mainly due to the small aperture size. The presented method has been applied for simple models and a vintage survey; it remains to be investigated how the method performs for longer offsets in a geologically more complex environment. Novel applications of residual migrations have been performed as one to one mappings; it has been demonstrated how summation approaches can be applied kinematically, where numerical implementations have still to be accomplished.

\section*{Acknowledgments} 
The author wishes to thank Dr. Sven Treitel for long standing advice and encouragement and Wintershall-DEA for permission to use the seismic survey. The Seismic Unix package has been used as a development platform for the seismic processing and for most of the graphical displays.

\section{references}
Adler, F. ,2002, Kirchhoff image propagation, Geophysics, 67, 126-134.
Al-Chalabi, M., 2014, Principles of seismic velocities and time-to-depth conversion: EAGE.\\
Bednar J., 2005, A brief history of seismic migration, Geophysics, 70, 3MJ-20MJ.\\
Born, M., 1985, Optik, 3rd edition (reprint), Springer.\\
Bording, P., Lines, L., 1997, Seismic modeling and imaging with the complete wave equation, SEG\\
Chauris, H., Noblez, M., Lambare, G, and P. Podvinz, 2002, Migration velocity analysis from locally coherent events in 2-D laterally heterogeneous media, Part I: Theoretical aspects, Geophysics, 67, 1202–1212\\
Deregowski, S. and Rocca, F., 1981, Geometrical optics and wave theory of constant offset sections in layered median: Geophysical Prospecting, 29, 374-406.\\
Etgen, J., 1998,  V(z) F-K prestack migration of common-offset common-azimuth data volumes, Part I: Theory,  SEG 68th international meeting, Expanded Abstracts, 1835-1838.\\
Fomel, S., 2014, Recent advances in time domain seismic imaging, SEG 84th international meeting, Expanded Abstracts, 4400-4404.\\
Gelchinsky, B. Y., 1961, Expressions for the spreading function: Problems of the Dynamic Theory of Seismic Wave Propagation, 5, 47–53, (in Russian).\\
Guillaume, P., Lambare, G., Leblanc, O. et alii ,2008, Kinematics invariants: an efficient and flexible approach for velocity model building. 78th SEG Annual International Meeting, Expanded Abstracts.\\
agedoorn, J., 1954, A process of seismic reflection interpretation: Geophysical Prospecting, 2, 85-127.\\
Hubral, P., 1977, Time migration - Some ray theoretical aspects: Geophys. Prosp., 25, 738-745.\\
Hubral, P. and Krey, T., 1980, Interval velocities from seismic refection time measurements: SEG.\\
Hubral, P., Schleicher, J. and M. Tygel, 1996, A unified approach to 3-D seismic reflection imaging,Part I: Basic concepts, 1996, Geophysics, 61,742-758\\
Iversen, M. Tygel, B. Ursin and M. de Hoop, 2012, Kinematic time migration and demigration of reflections in pre-stack seismic data: Geophys. J. Int., 189, 1635–1666.\\ 
Lambaré, G., P. Herrmann, P. Guillaume, S. Zimine, S. Wolfarth, O. Hermant, and S. Butt, 2007. From time to depth imaging with ‘Beyond Dix’, First Break, 25, 71-76.\\
Lambare, G., Herrmann, P., Toure, J., Suaudeau, E. and D. Lecerf, 2008, Computation of kinematic attributes for prestack time migration, SEG, 78th annual meeting, Expanded Abstracts, 2402-2406.\\
Lindsey J.P. 1989. The Fresnel zone and its interpretative significance. The Leading Edge, 8, 33-39.\\
Messud, J., P. Guillaume and G. Lambaré, 2017.  Estimating Structural Uncertainties in Seismic Images Using Equi-probable Tomographic Model, 79th EAGE Conference and Exhibition.\\
Montel, J.-P., N. Deladerriere, P. Guillaume, G. Lambaré, T. Prescott, J.-P. Touré, Y. Traonmilin, X. Zhang, 2009. Kinematic Invariants Describing Locally Coherent Events: An Efficient and Flexible Approach to Non-Linear Tomography, Extended Abstracts, 71st EAGE Conference and exhibition.\\
Poliannikov, O. and A. Malcolm, 2016: The effect of velocity uncertainty on migrated reflectors: Improvements from relative-depth imaging: Geophysics, 81, S21–S29.\\
Reshef, M. and D. Kosloff, 1986, Migration of common-shot gathers: Geophysics, 51, 324-331.\\
Robein, E., 2010, Seismic imaging: a review of  the techniques , their merits and limitations: EAGE Publications.\\ 
Rothman, D., Levin, S. and F. Rocca, 1985. Residual migration: Applications and limitations Geophysics, 50, 110-126.\\
Schneider, J., 1989, Specular prestack depth migration, 51 st conference and Exhibition, EAGE, preprint.\\
Schneider, J. 2011, Higher order residual moveout in horizon analysis, 73rd EAGE conference and Exhibition, EAGE, Extended Abstracts.\\
Schneider, J., 2014, Aspects of residual moveout after downward continuation: 76th conference, EAGE, Extended Abstracts.\\
Schneider, 2019, Horizon oriented residual prestack migration to zero offset, 89th SEG Annual International Meeting, Expanded Abstracts.\\
Sheriff R.E., 1980. Nomogram for Fresnel zone calculation. Geophysics 45,968-972.\\
Stolt, R., 1996, A prestack residual time migration operator, Geophysics, 61, 605-607.\\
Sun, J., 1996, The relationship between the first Fresnel zone and the normalized geometrical spreading factor, Geophysical Prospecting, 44, 351-374.\\
Tabti, H., Gelius, L., and T. Hellmann, 2004, Fresnel aperture prestack depth migration: First Break, 22, 39-46.\\
Vinje, V., Iversen, E. and H. Gjoeystdal, 1993, Traveltime and amplitude estimation using wavefront construction, Geophysics, 58,1157-1166.\\
Whitcombe, D., 1994, Fast model building using demigration and single-step ray migration: Geophysics, 59, 439–449.\\
Yilmaz, O, 2001, Seismic Data Analysis: SEG.\\
 
\appendices
\section{A zero offset Approximation for the Determination of PSTM Velocities after RMO analyses}

The lower figure in Fig. 1 is deduced from the upper part of Fig. 1 for the case of small receiver offsets for an isotropic PSTM. It is further assumed that the migration velocities around $P_{T0}$ do not vary significantly and the RMO determined for a particular source receiver offset is small, i.e. a locally homogeneous model is considered. The intersection point $F_M$ is determined from the emergence angles $\epsilon_S, \epsilon_G$  and the migration velocity $v_F$ at $P_{T0}$. It is then possible to apply zero offset considerations (Schneider, 2014): $F_M$ can be considered as a focus point of the downward continuation of the reflected event, the minimal distance $R_M = P_{T0}F_M$ between $F_M$ and $g_0$ as the residual radius of curvature induced by $v_F$. A direct estimate of the new migration velocity $v_N$ can be obtained from Gelchinskys law (Gel’chinsky, 1961) by requiring that $R_M$ vanishes:
\begin{equation}
\label{rdc}
 \Delta v_N^2=\frac{v_m^2R_I}{(T_0v_m+R_Itan^2\epsilon_I)cos^2\epsilon_I}
\end{equation}      
                                                     
with $R_I=P_{T0}Q_M+R_M=T_0v_M+R_M$, and the one way zero offset time $T_0$.\\
It has to be emphasized that the new migration velocity is estimated without knowledge of the true depth model.  For small offsets $v_N$ is optimal for a PSTM: in Fig. 1 for the case that $F_M$ and $P_{T0}$ coincide it is clear the aplanat also touches $g_0$. For small offsets h: $\epsilon_s-\epsilon_I=\epsilon_I-\epsilon_g+O[h^2]$ (Hubral and Krey, 1980, Appendix D), i.e. at $P_{T0}$ the event migrated with the new velocity $v_N$  will have the same slope as $g_0$.\\ 
A third application for the depth model in Fig. 3 is used to demonstrate the applicability of equation 1: for the dipping horizon a first PSTM was computed, where the maximum offset was confined to 2400m and migration velocities $v_{SAM}$ were computed along image rays (Hubral and Krey, 1980). After an RMO analysis the velocities $v_N$ were obtained from equation 1 (Fig. 5, magenta). The semblance coefficients in Fig. 19 demonstrate that there is hardly any residual moveout.\\
\\
Remarks:\\
\begin{itemize}
\item The new migration velocity has been derived from a downward continuation approach, whereas the application of the PSTM has been considered for common offset sections, e.g. according to the double square root equation (e.g. Lambare et alii, 2008). The derivation is applicable to the latter case because after a new migration with migration velocity $v_N$ the image of $F_m$ in Fig. 1 will be a focal point on the migrated image of $g_0$ by construction: the migrated times will be equal for zero offset and all small offsets events considered.   
\item The importance for practical applications should not be overemphasized: the values in magenta in Fig. 7 show that there is a distinct drop in semblance if larger offset traces are considered. Possible applications can be imagined in the reprocessing of vintage surveys or in using $v_N$ as an initial guess for a more eloborate parameter estimation.
\item It should be noted that there exist different possibilities to compute $v_N$:\\ 
If a depth model $v_F$ is available as in Fig. 2 $v_N$ can be estimated by requiring that the transmitted curvature $R_I$ (equation 1) of a zero offset ray raised from $P_{Z0}$ (Fig. 2) vanishes after downward continuation at the time migrated zero offset position $P_{T0}$. This task can be accomplished iteratively.\\
Alternatively, an estimate of $R_I$ can be obtained from stacking velocities. However, considering the similarity of $v_N$ and $v_{SAM}$ in Fig. 5, it is to be expected that the velocity bias between stacking and normal moveout velocities (Hubral and Krey, 1980, Al’Chalabi 2016) will impair the results, even for the short offset case considered.
\end{itemize}

\begin{figure}%[hbt]
  \captionsetup{justification=justified, singlelinecheck=off}
\centering
\includegraphics[width=0.45\textwidth]{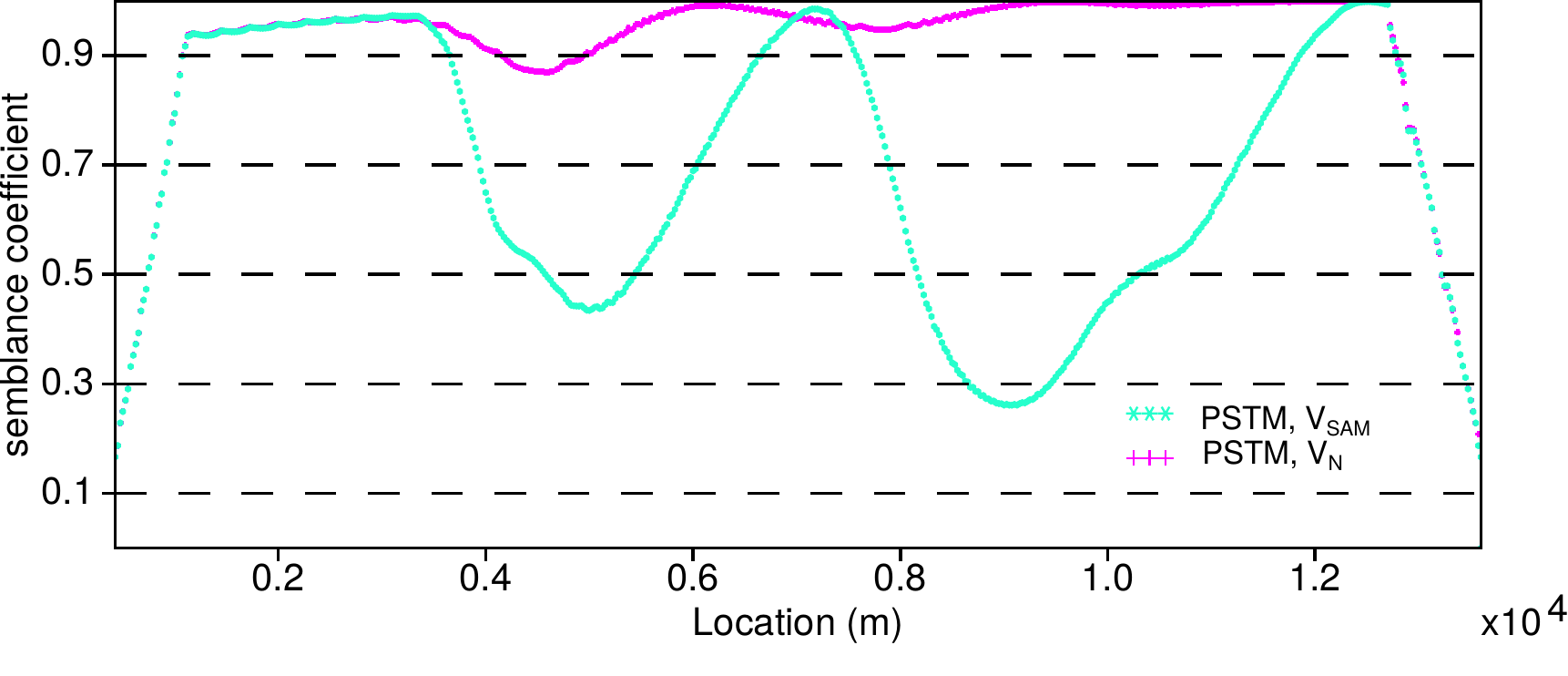}
\caption{Semblance coefficients for PSTM}
\label{fig:fig19}
\end{figure}

\end{document}